\documentclass[runningheads]{llncs}
\usepackage{booktabs}     
\usepackage[symbol]{footmisc}
\usepackage{xcolor,colortbl}
\usepackage{multirow}
\usepackage{tikz}
\usepackage{float}        
\usepackage{balance}
\usepackage{comment}      
\usepackage{caption}
\usepackage{array} 
\usepackage{subcaption}
\usepackage[table]{xcolor}
\usepackage{multirow}
\usepackage{changepage}
\usepackage{amsmath}
\usepackage{hyperref}

\title{A Comparative Study of Student Perspectives on Technical Writing Feedback Quality: Evaluating LLMs, SLMs, and Humans in Computer Science Topics}

\titlerunning{Comparative Study of Student Perspectives on Writing Feedback}

\author{%
Suqing Liu*\inst{1}\orcidID{0009-0004-5558-256X}\thanks{ These authors contributed equally.} \hfill \and
Runlong Ye*\inst{2}\orcidID{0000-0003-1064-2333} \and
Christopher Eaton\inst{3}\orcidID{0009-0002-2091-6625} \and
Bogdan Simion\inst{4}\orcidID{0000-0002-2554-8705} \and
Michael Liut\inst{4}\orcidID{0000-0003-2965-5302}}
\authorrunning{S. Liu et al.}
\institute{McMaster University, Hamilton, Ontario, Canada\\
\email{liu2684@mcmaster.ca} \and
Department of Computer Science, University of Toronto, Toronto, Ontario, Canada\\
\email{harryye@cs.toronto.edu} \and
Research Institute for the Study of University Pedagogy, University of Toronto Mississauga, Mississauga, Ontario, Canada\\
\email{chris.eaton@utoronto.ca} \and
Department of Mathematical and Computational Sciences, University of Toronto Mississauga, Mississauga, Ontario, Canada\\
\email{bogdan@cs.toronto.edu, michael.liut@utoronto.ca}}

\begin{document} 

\maketitle

\begin{abstract}


To address the scalability of feedback in computer science while mitigating the privacy and cost limitations of commercial Large Language Models (LLMs), this study evaluates a locally hosted Small Language Model (SLM). We compared a locally deployed 4-bit Llama-3.1-8B SLM, GPT-4o, and human feedback across introductory programming ($N=176$), operating systems ($N=80$), and a writing seminar ($N=7$). Mixed-methods analysis of student perceptions showed that the locally deployed SLM received the highest overall-quality ratings in two rubric-bound technical-writing contexts, while human feedback received the highest ratings in a small exploratory high-context writing pilot. These findings suggest that perceived feedback quality depends on the alignment between feedback form and assignment context, rather than model size alone. Local deployment reduced third-party data transfer and avoided per-token vendor fees after setup. Together, these results support a tiered feedback workflow in which local AI expands access to low-stakes, rubric-aligned first-pass feedback while human instructors focus their expertise on rhetorical nuance, conceptual judgment, and high-stakes assessment.


\end{abstract}

\section{Introduction}

Feedback enables students to understand their progress and identify areas that require improvement \cite{evans2013making,hattie2007power}. In education, both in the computing and writing/communication domains, where complex concepts and technical proficiency must be acquired, timely and constructive feedback is integral to student development \cite{bonsu2021influence,10.1145/3446871.3469768}. Traditionally, formative feedback in such contexts is often provided by Teaching Assistants, who guide learners through assignments, projects, and exercises. Although the quality and consistency of TA-generated feedback can vary \cite{cook}, students perceive it as a valuable and supportive resource \cite{kristiansen}.

As artificial intelligence (AI) technologies continue to evolve, recent work has proposed leveraging LLMs such as GPT \cite{openai2024gpt4technicalreport} or Llama \cite{grattafiori2024llama} as scalable, automated feedback generation tools \cite{kazemitabaar2024codeaid,pădurean2025humanizingautomatedprogrammingfeedback,2023Petersen-ITiCSEWG,zhang2022automated}. These AI-driven approaches hold the promise of delivering timely and consistent feedback at scale. Other work shows the impact of SLMs, lightweight models optimized for efficiency and accessibility, that have been used in place of LLMs \cite{a} as a time-efficient and cost-effective alternative. 


This research aims to address a gap in understanding how AI-generated feedback compares to human-provided alternatives by examining student perceptions of feedback quality and contextual suitability. Specifically, it evaluates the potential for LLMs and SLMs to deliver clarity, specificity, and actionable insights and to meet the nuanced requirements within computing education. Our key research questions (RQs) are:
\begin{enumerate}
    \item[] \textbf{RQ1}: How does students' perception of feedback quality differ between humans and language models? Furthermore, can students discern if the feedback is human\footnote{We define a \textit{human} as either a Teaching Assistant or Course Instructor.} or machine-generated?
    \item[] \textbf{RQ2}: How do local SLMs, commercial LLMs, and human instructors compare in terms of operational efficiency and scalability when providing technical feedback?
\end{enumerate}

\section{Related Work}
\subsection{Pedagogical Foundations of Effective Feedback}
Hattie and Timperley~\cite{hattie2007power} found that students place a premium on feedback that is personalized and actionable, enabling targeted improvements and fostering engagement. In computer science education, Fisk et al.~\cite{fisk2022automating} observed that automated personalized feedback on code significantly improved persistence, while Liu et al.~\cite{liu2024generating} found that timely, individualized feedback enhanced conceptual understanding during writing-to-learn activities. Similarly, Suraworachet et al.~\cite{suraworachet2023impact} showed that combining human and analytics-based feedback improved student engagement in reflective writing tasks. This evidence aligns with Watling's \cite {watling2019giving} emphasis on constructive, actionable, and iterative guidance. Such approaches ensure that students not only receive critiques, but can also integrate changes that lead to learning gains.

\subsection{Cognitive Load and Scaffolding}
The efficacy of such personalized guidance can be understood through the dual theoretical lenses of Cognitive Load Theory (CLT)\cite{sweller1988cognitive} and instructional scaffolding \cite{wood1976role}. CLT posits that working memory is limited; therefore, effective educational interventions must minimize extraneous cognitive load, such as poorly formatted or unstructured text, so learners can allocate mental resources to germane, schema-building tasks \cite{sweller1998cognitive}. For novice learners, rigid instructional scaffolding is essential to manage this cognitive load by breaking complex logic into manageable, step-by-step components. However, as learner expertise increases, the ``expertise reversal effect'' dictates that highly structured scaffolding becomes redundant, and learners instead require nuanced, high-level conceptual guidance \cite{kalyuga2009expertise}. This theoretical dichotomy directly informs our investigation into whether the structured formatting of AI feedback acts as an effective cognitive scaffold, and how students' perceptions of machine versus human quality shift alongside their developing expertise.

\subsection{AI-Driven Feedback Systems and SLMs}
The use of LLMs as virtual teaching assistants has been explored, demonstrating their effectiveness in providing clear, engaging feedback in programming courses, though human supervision remains crucial~\cite{10.1145/3649217.3653574,10.1145/3626252.3630789}. Automated Feedback Systems (AFSs), driven by advancements in NLP and machine learning, offer scalable solutions for generating personalized feedback on writing~\cite{huawei}. Son et al.~\cite{Son} highlight the growing adoption of AI-driven feedback for grammar, structure, and coherence in language learning, effectively reducing instructor workload through the integration of automated grading and feedback systems (reducing instructor grading hours from 15 to 9.75 per week). Similarly, in computer science education, studies by Wei~\cite{10.3389/fpsyg.2023.1261955} demonstrates how AI systems can provide error-specific feedback. Beyond individual courses, AFSs have been linked to improved engagement and retention in online education \cite{villegas2024application}. Students receiving AI-generated feedback were more likely to complete assignments and show deeper engagement, suggesting that real-time, automated guidance can be a powerful motivator.

While prior research highlights the benefits of AI-generated feedback, comparisons against human feedback from the student perspective remain sparse. For instance, although Guo and Wang \cite{guo2024resist} found complementary affordances between ChatGPT and teacher feedback in an EFL context, their evaluation centered entirely on teacher perceptions (N=5). How computing students perceive this dynamic, particularly when utilizing locally deployable SLMs for data privacy and scalability \cite{a}, has not been thoroughly explored. We address these gaps by conducting a comprehensive, mixed-methods comparison of the perceived feedback quality generated by LLMs, SLMs, and humans across diverse instructional contexts.

\section{Methodology}

This study was conducted at a large, publicly funded, research-focused university in North America. We selected three distinct undergraduate courses to evaluate students' reception of feedback efficacy across varying levels of domain complexity and technical depth. All data collection was approved by the Institutional Review Board (IRB), and students provided informed consent for their data to be used for analysis.

\subsection{Course Context and Participants}

\textbf{Introduction to Computer Science (CS2, $n=176$):} This large first-year course (694 students enrolled) focuses on software design fundamentals in Python, including object-oriented programming, recursion, and complexity analysis. For this study, students submitted a class hierarchy implementation accompanied by a technical design document. The document was assessed on structure, writing mechanics, and audience awareness. There were 190 unique survey submissions from this class, 176 of which passed the survey's attention check.

\textbf{Operating Systems (OS, $n=80$):} This specialized third-year course (143 students enrolled) covers system-level concepts such as concurrency, memory management, and file systems. The study focused on a synchronization assignment where students were required to submit two technical artifacts: a bug report detailing a race condition and a README discussing starvation risks in their design. These tasks emphasized precise technical reasoning and concise communication. There were 83 unique survey submissions from this class, 80 of which passed the survey's attention check.

\textbf{Writing on AI Topics ($n=7$):} This third-year seminar course (13 students enrolled) explores the intersection of artificial intelligence and communication design. Students produced an eight-page professional proposal aimed at industry decision-makers. The assignment required the integration of peer-reviewed research and multimodal elements to argue for AI-driven workflow improvements. There were seven unique survey submissions from this class, all of which passed the survey's attention check.

\subsection{Feedback Generation System}
Our RAG pipeline builds on insights from our prior work \cite{10.1145/3649217.3653554,a}. In particular, we adopt and adapt the recommendations around vector database design, retrieval accuracy, and hallucination mitigation, tailoring the approach to support rubric-aligned feedback rather than open-ended student assistance.

Two distinct model architectures were employed:
\begin{enumerate}
    \item \textbf{Large Language Model (LLM):} \texttt{GPT-4o}, selected to represent state-of-the-art cloud-based reasoning.
    
    \item \textbf{Small Language Model (SLM):} Meta's \texttt{Llama-3.1-8B}\footnote{Specifically, we used
    \texttt{unsloth/Meta-Llama-3.1-8B-Instruct-bnb-4bit}, available on \href{https://huggingface.co/unsloth/Meta-Llama-3.1-8B-Instruct-bnb-4bit}{Hugging Face}.}, deployed locally using 4-bit quantization and \verb|torch.compile()| optimization. This model served as an institutionally hosted alternative designed to reduce third-party data transfer and recurring API costs.\. See Appendix for complete technical specifications of local SLM deployment.
\end{enumerate}

Both models used identical system prompts (see Appendix) designed to generate constructive, criterion-based feedback without assigning numeric grades.

\subsection{Data Collection \& Survey Design}

We employed a blinded, within-subjects experimental design. Students accessed a web interface displaying three anonymized feedback responses for their submission: one from the LLM, one from the SLM, and one from a human evaluator (TA or Course Instructor). The display order was randomized.

The survey consisted of three components:

\textbf{1. Preference \& Rationale:} Participants selected their single most preferred feedback source and provided a qualitative justification (\emph{$Q_{Reason}$}).

\textbf{2. Quantitative Ratings:} Students evaluated each feedback source on a 7-point Likert scale (1=Lowest, 7=Highest) across six dimensions derived from feedback literature~\cite{hattie2007power,nicol2006formative}: \textit{Readability}, \textit{Detail}, \textit{Specificity}, \textit{Actionability}, \textit{Helpfulness}, and \textit{Overall Quality}. An attention check item was embedded to validate response quality.

\textbf{3. Source Discernment \& Reflection:} Students attempted to identify the origin of each feedback item (AI vs. Human) (\emph{$Q_{AIHuman}$}). Finally, two open-ended questions invited suggestions for system improvement (\emph{$Q_{Improve}$}) and broader reflections on the role of AI in assessment (\emph{$Q_{Discern}$}).

\subsection{Evaluation}

\paragraph{Quantitative Analysis.}
As the Likert data violated normality assumptions (Shapiro-Wilk, $p < .05$), we utilized non-parametric methods. Differences between feedback sources were assessed using the Friedman test for repeated measures. Significant omnibus results were followed by post-hoc Wilcoxon signed-rank tests, with p-values adjusted using the Holm-Bonferroni correction to control for family-wise error rates.

\paragraph{Qualitative Analysis.}
Responses to open-ended items ($Q_{Reason}$, $Q_{Improve}$, $Q_{Discern}$) were analyzed using a consensus-based thematic analysis~\cite{braun2006using}. Rather than relying on a rigid, deductive codebook, two researchers independently reviewed a subset of responses to generate an initial thematic framework, which was iteratively refined through collaborative discussion. This organic process focused on identifying the specific textual features, such as tone, structural specificity, and perceived correctness, that drove student preferences.

Following the methodological arguments of Hammer and Berland \cite{hammer2014confusing}, we prioritized negotiated agreement over statistical inter-rater reliability. In this paradigm, coding is not the independent categorization of isolated statements, but rather a reflexive process of developing a shared understanding of complex, subjective student data. Therefore, rigour was established through the dialogic resolution of discrepancies, the maintenance of a transparent audit trail, and the deliberate discussion of negative or disconfirming cases across varying levels of student expertise.

\section{Results}
\begin{figure}[]
    \centering
    \begin{subfigure}[t]{\linewidth}
        \centering
        \includegraphics[width=1\linewidth]{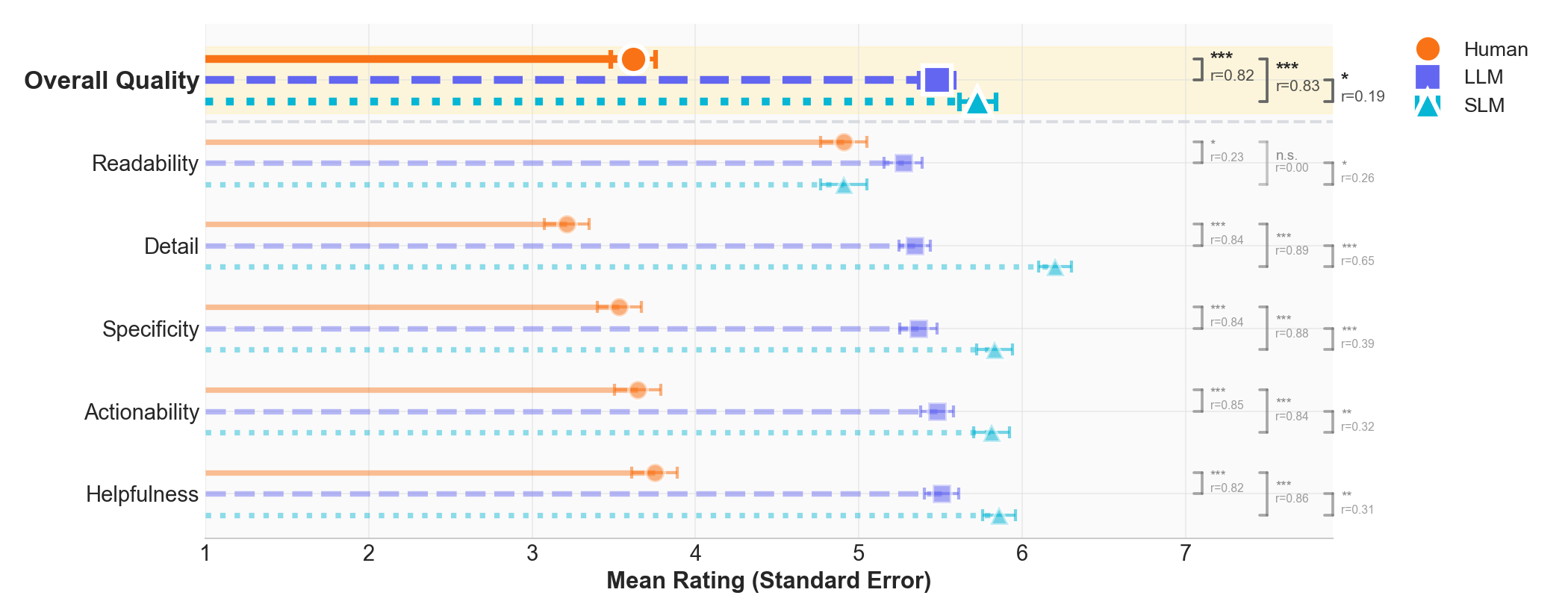}
        \caption{CS2 -- Introduction to Computer Science (N = 176)}
        \label{fig:cs2}
    \end{subfigure}
    \begin{subfigure}[t]{\linewidth}
        \centering
        \includegraphics[width=1\linewidth]{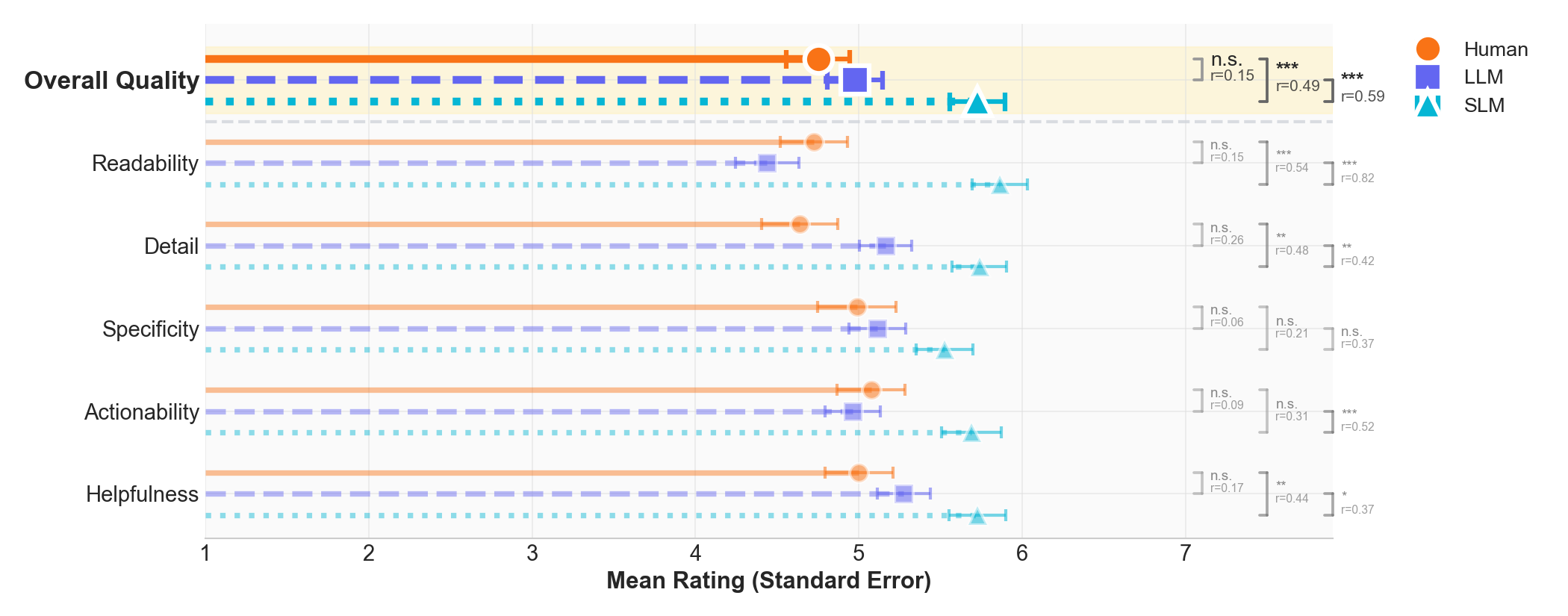}
        \caption{Operating Systems (N = 80)}
        \label{fig:os}
    \end{subfigure}
    \begin{subfigure}[t]{\linewidth}
        \centering
        \includegraphics[width=1\linewidth]{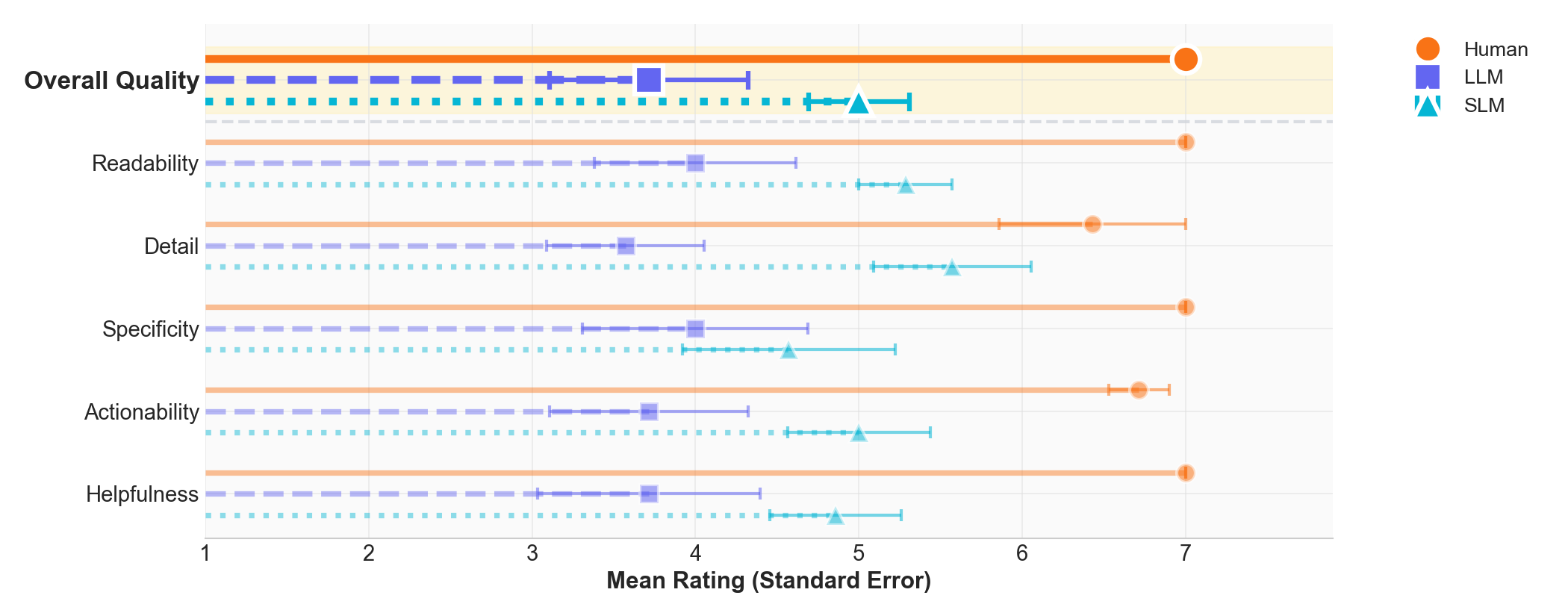}
        \caption{Writing on AI Topics (N = 7)}
        \label{fig:writing}
    \end{subfigure}
     \caption{ Student feedback ratings across three courses. Ratings are measured on a 1--7 Likert scale (x-axis: higher is better). \textbf{n.s.}: $p\geq.05$; \textbf{*}: $p<.05$; \textbf{**}: $p<.01$; \textbf{***}: $p<.001$ (Holm corrected). Effect sizes are reported as rank-biserial $r$.  Inferential statistics are omitted from (c) due to the pilot’s small sample size.}    
    \label{fig:combined_vertical}
\end{figure}

We present our quantitative results in \autoref{fig:combined_vertical}. The x-axis represents mean student ratings, where 7 indicates the best score and 1 the worst. The y-axis captures feedback dimensions, ranging from readability and level of detail to overall quality.

Below, we report on the student ratings on the various dimensions of feedback quality from LLMs, SLMs, and human.

\subsection{Quantitative Findings}

\paragraph{Statistical Power and Pilot Designation.}
Before evaluating specific feedback dimensions, we conducted a post-hoc power analysis to determine the statistical adequacy of our sample sizes for the Wilcoxon signed-rank tests. Assuming a medium effect size ($r = 0.3$) and $\alpha = .05$, the analysis confirmed that our largest cohort, the introductory CS2 course ($n = 176$), was highly powered at 97.8\%. The advanced Operating Systems course ($n = 80$) achieved 76.5\% power, which is marginally below the 80\% standard but sufficient for detecting moderate-to-large effects \cite{cohen2013statistical}. Expectedly, the Writing on AI Topics cohort ($n = 7$) achieved only 12.5\% power. Consequently, we excluded the writing cohort from any cross-course inferential generalizations and treated it as exploratory pilot data. 

\paragraph{General Trends in Technical Courses.} Across the two adequately powered technical courses (CS2 and OS), a distinct hierarchy in student preference emerged. The Small Language Model (SLM) consistently achieved the highest mean ratings for \textit{Overall Quality} and \textit{Detail}, considerably outperforming the Human baseline. While the commercial LLM was generally rated higher than Human feedback overall, it did not consistently match the \textit{Specificity} and \textit{Actionability} of the locally hosted SLM. However, the dynamics between the models shifted across assignment contexts, as detailed below.

\paragraph{Course-Specific Quality Analysis: CS2.} In the introductory CS2 course ($n = 176$), students consistently rated Human feedback lower than both AI sources. The mean Overall Quality for Human feedback was 3.62, significantly lower than that of the LLM ($M = 5.48, p < .001, r = 0.82$) and SLM ($M = 5.73, p < .001, r = 0.83$). Notably, while the SLM received the highest ratings for content-related metrics such as Detail ($M = 6.20$) and Helpfulness ($M = 5.86$), the generic LLM was rated highest for Readability ($M = 5.27$), significantly outperforming the SLM ($M = 4.91, p = .024, r = 0.26$) and Human feedback ($M = 4.91, p = .021, r = 0.23$).

\paragraph{Course-Specific Quality Analysis: Operating Systems.} In the advanced Operating Systems course ($n = 80$), the advantage of the generic LLM disappeared. There was no significant difference in Overall Quality ratings between the LLM ($M = 4.97$) and Human feedback ($M = 4.75, p = .28, r = 0.15$). However, the specialized SLM maintained its lead, achieving the highest ratings in the cohort for Overall Quality ($M = 5.72$), significantly outperforming both the Human ($p < .001, r = 0.49$) and LLM baselines ($p < .001, r = 0.59$). A key divergence occurred in Readability: unlike in the introductory course, OS students rated the SLM as the most readable source ($M = 5.86$), significantly higher than the LLM ($M = 4.44, p < .001, r = 0.82$) and Human feedback ($M = 4.72, p < .001, r = 0.54$).

\paragraph{Exploratory Pilot Data: Writing on AI Topics Quality Analysis.} Descriptive statistics for the Writing on AI Topics pilot ($n = 7$) show a trend reversal compared to the technical courses. Human feedback received a high rating for Overall Quality ($M = 7.00$), Descriptive Specificity ($M = 7.00$), and Helpfulness ($M = 7.00$). In contrast, the AI models received substantially lower descriptive ratings, with the SLM ($M = 5.00$) and LLM ($M = 3.71$) rated below human feedback. While the underpowered sample size prevents inferential generalization, the descriptive data suggest a possible domain-specific preference for human feedback in high-context rhetorical scenarios.

\begin{figure}[htbp]
    \centering
    \begin{subfigure}[b]{0.32\textwidth}
        \centering
        \includegraphics[width=\linewidth]{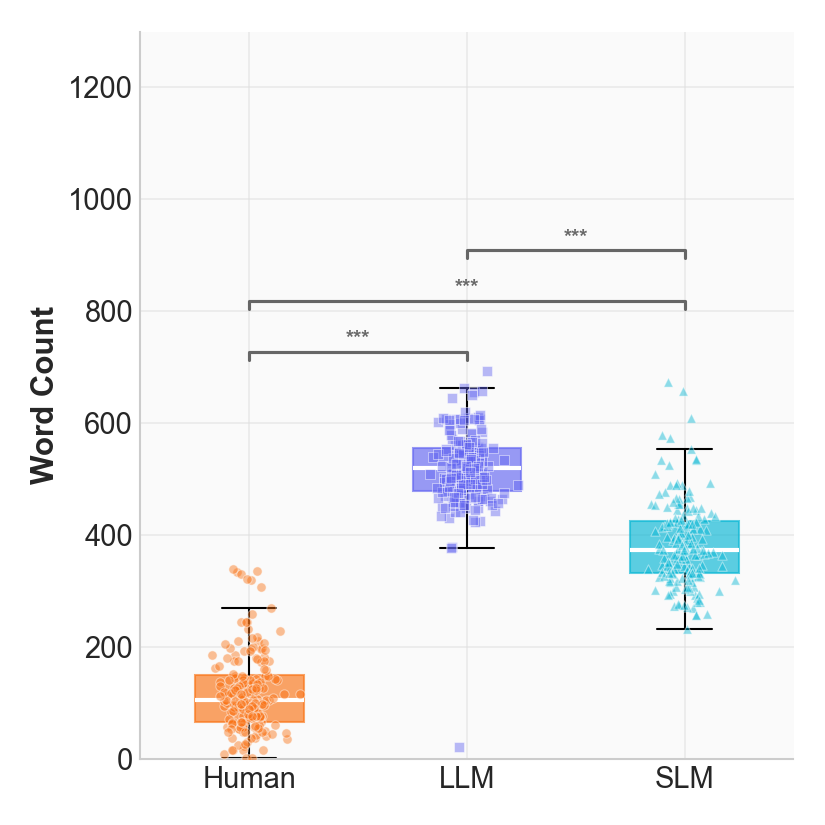}
        \caption{Response word count: CS2}
        \label{fig:sub1}
    \end{subfigure}
    \hfill
    \begin{subfigure}[b]{0.32\textwidth}
        \centering
        \includegraphics[width=\linewidth]{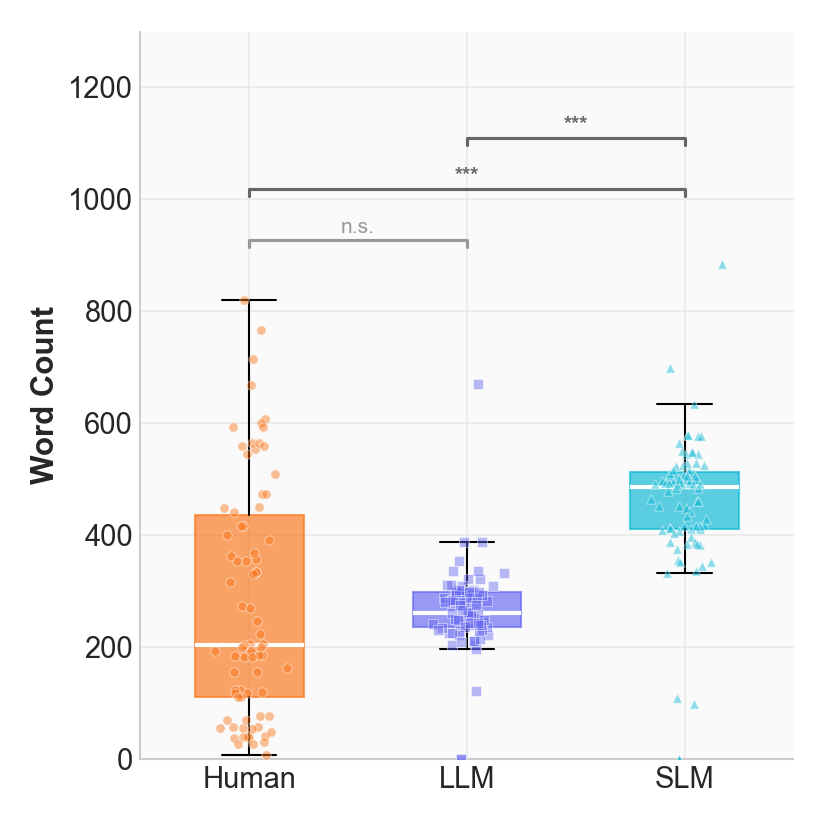}
        \caption{Response word count: OS}
        \label{fig:sub2}
    \end{subfigure}
    \hfill
    \begin{subfigure}[b]{0.32\textwidth}
        \centering
        \includegraphics[width=\linewidth]{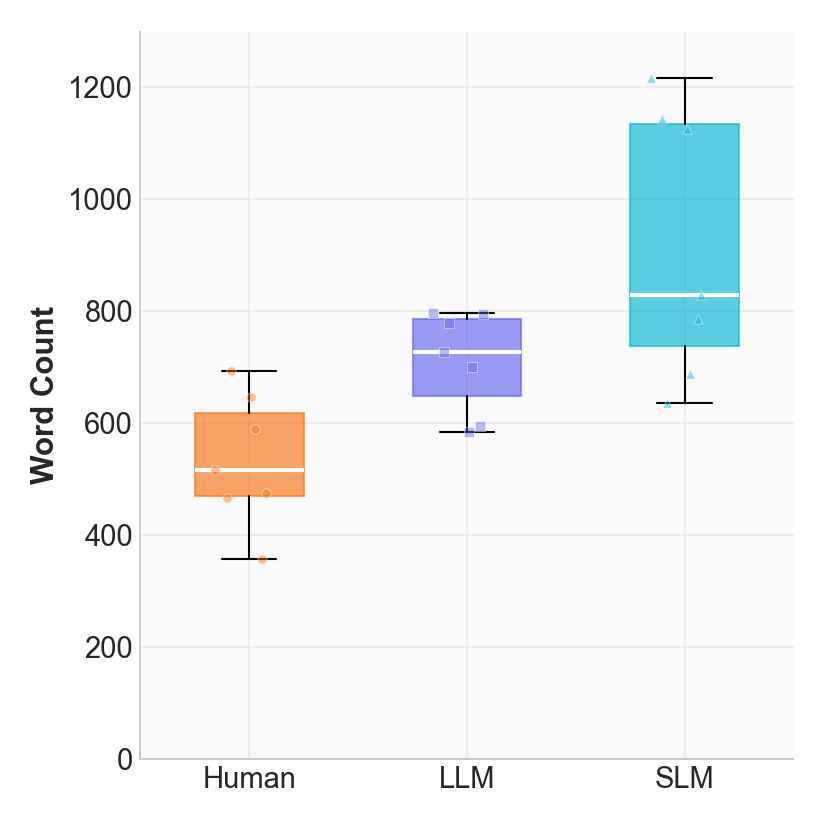}
         \caption{Response word count: Writing on AI Topics}
        \label{fig:sub3}
    \end{subfigure}
    \caption{Word count distributions of feedback generated by Human, LLM, and SLM across (a) CS2, (b) Operating Systems, and (c) Writing on AI Topics. Statistical significance is derived from post-hoc Wilcoxon signed-rank tests (\textbf{n.s.}: $p\geq.05$; \textbf{*}: $p<.05$; \textbf{**}: $p<.01$; \textbf{***}: $p<.001$, Holm-corrected). Inferential statistics are omitted from (c) due to the pilot's small sample size.}
    \label{fig:feedback_length}
\end{figure}

\paragraph{Feedback Length Analysis.}
To objectively compare the structural differences between the feedback sources, we analyzed the word count distributions across all three courses (Figure \ref{fig:feedback_length}). 

In the introductory CS2 course ($n = 176$), a Friedman test revealed significant differences in feedback length ($\chi^2 = 313.99, p < .001$). Post-hoc Wilcoxon signed-rank tests (Holm-corrected) indicated that both the LLM and SLM generated significantly more text than human instructors ($p < .001$, with large effect sizes of $r > 0.99$ for both). Furthermore, the LLM's feedback was significantly longer than the SLM's ($p < .001, r = 0.92$).

In the advanced Operating Systems course ($n = 80$), the length dynamics shifted ($\chi^2 = 65.75, p < .001$). Here, the SLM produced significantly longer feedback than both human instructors ($p < .001, r = 0.73$) and the LLM ($p < .001, r = 0.98$). Notably, there was no significant difference in word count between the human and LLM feedback ($p = .79$). Finally, descriptive data from the Writing pilot ($n=7$) indicated a similar trend where the SLM generated the highest volume of text, though inferential testing was omitted due to the underpowered sample size.


\begin{table}[t]
\centering
\scriptsize
\caption{Student attribution of feedback as AI or human, grouped by actual feedback source.}
\setlength{\tabcolsep}{4pt}
\renewcommand{\arraystretch}{0.95}
\begin{tabular*}{\textwidth}{@{\extracolsep{\fill}} l ccc ccc ccc}
\toprule
& \multicolumn{3}{c}{\textbf{Introductory CS2}} 
& \multicolumn{3}{c}{\textbf{Operating Systems}} 
& \multicolumn{3}{c}{\textbf{Writing on AI Topics}} \\
\cmidrule(lr){2-4} \cmidrule(lr){5-7} \cmidrule(lr){8-10}
\textbf{Actual Source} 
& \textbf{AI} & \textbf{Human} & \textbf{Acc.}
& \textbf{AI} & \textbf{Human} & \textbf{Acc.}
& \textbf{AI} & \textbf{Human} & \textbf{Acc.} \\
\midrule
LLM   & 61 & 115 & 34.66\% & 41 & 39 & 51.25\% & 7 & 0 & 100\% \\
SLM   & 93 & 83  & 52.84\% & 61 & 19 & 76.25\% & 7 & 0 & 100\% \\
Human & 51 & 125 & 71.02\% & 15 & 65 & 81.25\% & 0 & 7 & 100\% \\
\bottomrule
\end{tabular*}
\label{tab:AIvHuman}
\end{table}

\paragraph{Distinguishing AI from Human Feedback.} Discernment accuracy varied significantly by student experience and model sophistication (see~\autoref{tab:AIvHuman}). Advanced Operating Systems students (69.58\% overall accuracy) reliably distinguished Human (81.25\%) and SLM (76.25\%) authors, but found the commercial LLM nearly indistinguishable from human feedback (51.25\%). Novice CS2 students were even more susceptible to the AI's ``human-like'' quality, correctly identifying the LLM and SLM sources only 34.66\% and 52.84\% of the time, respectively.

These results suggest that while students in the AI topics writing course could accurately distinguish the language model feedback from the instructor’s feedback, OS students distinguished AI from human feedback more reliably, whereas CS2 students were less reliable overall and frequently misidentified the commercial LLM as human-generated. The Writing on AI Topics result should be treated as descriptive only as the cohort was very small.

\subsection{Qualitative Findings}

\paragraph{Formatting as Function: Students Prefer the SLM’s Rigid, Step-by-Step Structure for Technical Tasks.}
Contrary to the expectation that quantized models might degrade coherence, students in technical courses specifically favored the SLM for its rigid adherence to formatting. The local \texttt{Llama-3.1} model consistently generated feedback with explicit headers, bullet points, and distinct ``Reasoning'' sections (a byproduct of the system prompt). For CS2 and OS students, this structural rigidity mirrored technical documentation standards, making the feedback appear more rigorous than the conversational style of the human instructors. One student noted, \textit{``The layout [of the SLM] makes it easier to debug my logic step-by-step,''} indicating that for technical writing, formatting is a functional component of feedback quality.

\paragraph{Beyond Word Count: Structured Coverage as a Cue for Perceived Usefulness.}
The word-count analysis helps clarify that students were responding to more than length alone. In CS2, the LLM generated significantly longer feedback than the SLM, yet the SLM received the highest overall-quality rating. In Operating Systems, the SLM produced the longest feedback and was also rated highest. In the Writing on AI Topics pilot, the SLM again produced the most text descriptively, but human feedback received the highest ratings. Taken together, these patterns point away from a simple word-count explanation and toward the perceived usability of structured, visibly comprehensive feedback.

Qualitative responses support this interpretation. Some students described detailed AI feedback as evidence that their work had been carefully reviewed, with one student saying that it felt as though the AI had read every line. In the technical-writing contexts, this visible coverage appeared most valuable when paired with criterion-by-criterion organization, reasoning sections, and concrete suggestions. Thus, the SLM’s advantage is best understood as a combination of structured presentation, rubric coverage, and actionable detail rather than sheer response length.

\paragraph{Low-Stakes AI Feedback as a Perceived Space for Early Drafting.}
Students explicitly recognized the trade-off between the ``warmth'' of human feedback and the immediate utility of the local model. The perception of the SLM as a \textit{``helpful and cheap service''} highlights a shift in student expectations: they view the local model not as a teacher replacement, but as a high-availability tool for logic confirmation. Some students described AI feedback as a lower-stakes space for early drafting, where they could seek clarification without feeling judged by a human evaluator. One participant noted, \textit{``AI is friendlier than a human because I don't feel judged for bad drafts,''} suggesting that the ``cold'' tone of the local model acts as a privacy-preserving feature that encourages experimentation.

\paragraph{A Boundary Condition: The Deployed Local SLM Struggled with Rhetorical Nuance.}
While the SLM excelled at syntax and structure, qualitative responses revealed limitations of the deployed local SLM setup in high-context rhetorical tasks. In high-context scenarios (i.e., the Writing seminar), the SLM struggled to interpret editorial intent, often correcting stylistic choices as if they were objective errors (\textit{``The AI missed the point of what I was trying to argue''}). This limitation validates the quantitative drop in performance for the writing course; while the SLM can parse technical specifications in a prompt (RAG), the deployed local SLM setup appeared less responsive to rhetorical intent and contextual judgment in this assignment., showcasing a clear boundary for where human review remains especially valuable.

\section{Discussion}

\subsection{RQ1: Task--Feedback Fit Across Technical and Rhetorical Writing Contexts}

Our findings suggest a tiered utility for AI in technical education, where perceived feedback quality depends on the alignment between feedback form and assignment context rather than model parameter size alone. In the two rubric-bound technical-writing contexts, CS2 and Operating Systems, students rated the locally deployed SLM highest overall. Qualitative responses suggest that students valued the SLM’s criterion-by-criterion structure, explicit reasoning sections, and concrete suggestions. These features made the feedback easy to navigate and well aligned with assignments where students needed structured guidance on technical communication, documentation, and revision.

Viewed through the lens of Cognitive Load Theory \cite{sweller1988cognitive}, the SLM’s rigid formatting can be understood as useful scaffolding for parsing complex feedback. Explicit headings, rubric-aligned organization, and step-by-step suggestions may reduce the effort required to locate and act on feedback, especially in technical-writing tasks with clearly defined criteria. The qualitative responses also suggest that the SLM created a low-stakes space for early revision, where students could use automated feedback for clarification and logic confirmation before seeking higher-level human guidance.

The same pattern did not hold in the Writing on AI Topics pilot. In that small exploratory context, human feedback received the highest descriptive ratings, and students indicated that AI feedback sometimes missed rhetorical intent. This result highlights a productive boundary condition: structured local AI feedback appears well suited to low-stakes, rubric-aligned technical writing, while human feedback remains especially valuable when feedback depends on audience, argument, purpose, and contextual judgment.

We therefore interpret the results as evidence of task--feedback fit rather than student expertise alone. Operating Systems students were more advanced than CS2 students but still rated the SLM highest overall, suggesting that the assignment’s structure and feedback needs mattered more than course level by itself. This interpretation positions local SLMs as useful partners for scalable technical-writing support while preserving a central role for human instructors in contexts that require rhetorical and conceptual judgment.

\paragraph{Perception and Detection.}

At the aggregate course level, perceived authorship and perceived quality were partly dissociated. In Operating Systems, 76.25\% of students correctly identified the SLM feedback as AI-generated, yet the SLM still received the highest overall-quality rating. In CS2, GPT-4o was most frequently mistaken for human feedback, but it did not receive the highest overall rating. These patterns suggest that feedback did not need to appear human to be perceived as useful.

This strengthens the practical case for local AI as a distinct educational tool rather than a simulated instructor. Students did not need AI feedback to perfectly imitate human feedback in order to value it. Instead, the results suggest that students responded to the feedback’s usability, structure, and relevance to revision. This finding supports a design goal in which AI feedback is transparent, task-aligned, and useful, rather than primarily optimized to appear human.

\subsection{RQ2: Efficiency, Sustainability, and Institutional Control}

The deployment of automated feedback pipelines highlights a substantial contrast in resource utilization between human and AI workflows. Human graders in the operating-systems course invested approximately 40 hours, including grading and training, to generate feedback, while the instructor in the writing seminar spent 4.5 hours. In contrast, our automated pipeline processed the entire operating-systems cohort in 1.8 hours and the writing seminar in under 10 minutes. This result shows the practical value of automated feedback for routine, batch-processed formative feedback, where rapid turnaround and consistent rubric coverage are central instructional needs.

The comparison between cloud and local AI further highlights an important institutional trade-off. The commercial LLM (\texttt{GPT-4o}), while offering higher inference throughput ($\sim$26 tokens/s), accumulated approximately \$500 in API usage fees across two rounds of feedback. This variable cost structure can become a planning challenge for large-enrolment courses that require repeated rounds of feedback.

The locally hosted SLM offered a complementary deployment profile. By applying 4-bit quantization to the \texttt{Llama-3.1-8B} model, we reduced the memory footprint to fit comfortably within the VRAM limits of a standard GPU ($\geq$12\,GB). Although the local inference speed was lower ($\sim$15 tokens/s), this latency is acceptable in an asynchronous batch-processing context. Once deployed, the local SLM avoided per-token vendor charges and allowed student submissions to be processed within institutionally managed infrastructure.

These results position local SLMs as a promising architecture for scalable, institutionally controlled feedback in courses where feedback can be generated asynchronously and aligned to explicit rubrics \cite{a}. Rather than replacing either commercial LLMs or human instructors, the local model expands the design space for feedback systems by offering a practical middle ground: more scalable than human-only workflows, more institutionally controlled than cloud-only workflows, and sufficiently useful for structured technical-writing feedback.

\section{Limitations and Future Work}

Our results are constrained by the context window limitations of local models, which occasionally resulted in generic feedback when the error required deep knowledge of assignment constraints. Additionally, our sample skewed toward upper-year students, and generalizability to non-technical domains remains untested. Methodologically, as indicated by our post-hoc power analysis, the advanced and writing-intensive cohorts possessed limited statistical power. Consequently, findings from the writing seminar must be treated as exploratory pilot data rather than inferential generalizations.

Crucially, our evaluation is bounded by student self-reports and ratings. While these metrics provide valuable insights into user acceptance, readability, and perceived utility, they do not substantiate claims of objective educational effectiveness. As noted in prior literature, students may prefer feedback that reduces cognitive load rather than feedback that maximizes learning. Future work must incorporate pre- and post-learning assessments, learning analytics (e.g., LMS logs), and student revision tracking to determine whether highly-rated SLM feedback translates into measurable pedagogical outcomes. 

On the technical side, our evaluation focused on a single commercial LLM and a locally hosted SLM. Refining the system prompt for AI models to strictly enforce a rigid structural template or evaluating state-of-the-art frontier models could alter the preference gap. Furthermore, our pipeline utilized a fixed 4-bit quantization; future technical evaluations should include a sensitivity analysis across various depths (e.g., 2-bit, 8-bit, 16-bit) to map the precise trade-off between memory footprint and feedback quality. To efficiently scale these comparative studies, future work could leverage unified API routing tools, such as OpenRouter, to query and benchmark feedback quality across a wide range of open-source and proprietary models without increasing infrastructure overhead.

\section{Conclusion}

To address the privacy, cost, and scalability bottlenecks of commercial language models in computing education, this study evaluated students’ perceived feedback quality for a locally hosted, quantized Small Language Model (SLM), GPT-4o, and human feedback. Through a mixed-methods, within-subjects evaluation across introductory programming (CS2), operating systems (OS), and an exploratory writing pilot, we showed that model size does not strictly determine perceived feedback quality. Across two rubric-bound technical-writing contexts, students rated the locally deployed SLM highest overall, particularly for dimensions such as detail, actionability, and helpfulness. In contrast, human feedback received the highest descriptive ratings in a small exploratory high-context writing pilot.

Together, these findings support a task-sensitive approach to AI feedback in computing education. Local SLMs can expand access to timely, structured, rubric-aligned feedback in technical-writing contexts, while human instructors remain essential for rhetorical nuance, conceptual judgment, and high-stakes assessment. Operationally, the local SLM reduced routine feedback-generation time, avoided per-token vendor fees after deployment, and allowed feedback generation to occur within institutionally managed infrastructure. The resulting framework is not a competition between local AI, cloud AI, and human instructors, but an opportunity to coordinate their strengths. By using local SLMs for scalable first-pass feedback and preserving human attention for the forms of judgment students value most, institutions can build feedback systems that are more timely, sustainable, and pedagogically responsive.

\begin{credits}
\subsubsection{\ackname}
We would like to thank the Learning \& Education Advancement Fund (LEAF) from the Office of the Vice-Provost, Innovations in Undergraduate Education, University of Toronto, and the Natural Sciences and Engineering Research Council of Canada (NSERC) Discovery Grant (\#RGPIN-2024-04348). We would also like to thank \href{https://www.cas.mcmaster.ca/~deza/}{Dr. Antoine Deza} for his continued support, guidance, and mentorship. 
\end{credits}

\bibliographystyle{splncs04}
\bibliography{sn-bibliography}

@article{hattie2007power,
  title={The power of feedback},
  author={Hattie, John and Timperley, Helen},
  journal={Review of educational research},
  volume={77},
  number={1},
  pages={81--112},
  year={2007},
  publisher={Sage Publications Sage CA: Thousand Oaks, CA}
}

@inproceedings{kristiansen,
  title={Feedback on student programming assignments: Teaching assistants vs automated assessment tool},
  author={Kristiansen, Nynne Grauslund and Nicolajsen, Sebastian Mateos and Brabrand, Claus},
  booktitle={Proceedings of the 23rd Koli Calling International Conference on Computing Education Research},
  pages={1--10},
  year={2023}
}

@inproceedings{cook,
  title={Improving TA feedback on in-class coding assignments for introductory computer science},
  author={Cook, Amy and Phan, Vinhthuy and Windsor, Alistair},
  booktitle={Proceedings of the 27th ACM Conference on on Innovation and Technology in Computer Science Education Vol. 1},
  pages={421--427},
  year={2022}
}

@article{openai2024gpt4technicalreport,
  title={Gpt-4 technical report},
  author={Achiam, Josh and Adler, Steven and Agarwal, Sandhini and Ahmad, Lama and Akkaya, Ilge and Aleman, Florencia Leoni and Almeida, Diogo and Altenschmidt, Janko and Altman, Sam and Anadkat, Shyamal and others},
  journal={arXiv preprint arXiv:2303.08774},
  year={2023}
}

@inproceedings{zhang2022automated,
  title={Automated feedback generation for competition-level code},
  author={Zhang, Jialu and Li, De and Kolesar, John Charles and Shi, Hanyuan and Piskac, Ruzica},
  booktitle={Proceedings of the 37th IEEE/ACM international conference on automated software engineering},
  pages={1--13},
  year={2022}
}

@article{huawei,
  title={A systematic review of AI-based automated written feedback research},
  author={Shi, Huawei and Aryadoust, Vahid},
  journal={ReCALL},
  volume={36},
  number={2},
  pages={187--209},
  year={2024},
  publisher={Cambridge University Press}
}

@article{Son,
  title={Artificial intelligence technologies and applications for language learning and teaching},
  author={Son, Jeong-Bae and Ru{\v{z}}i{\'c}, Natasha Kathleen and Philpott, Andrew},
  journal={Journal of China computer-assisted language learning},
  volume={5},
  number={1},
  pages={94--112},
  year={2025},
  publisher={De Gruyter}
}

@ARTICLE{10.3389/fpsyg.2023.1261955,
  title={Artificial intelligence in language instruction: impact on English learning achievement, L2 motivation, and self-regulated learning},
  author={Wei, Ling},
  journal={Frontiers in psychology},
  volume={14},
  pages={1261955},
  year={2023},
  publisher={Frontiers Media SA}
}

@article{villegas2024application,
  title={Application of Artificial Intelligence in Online Education: Influence of Student Participation on Academic Retention in Virtual Courses},
  author={Villegas-Ch, William and Garc{\'\i}a-Ortiz, Joselin and S{\'a}nchez-Viteri, Santiago},
  journal={IEEE Access},
  year={2024},
  publisher={IEEE}
}

@article{watling2019giving,
  title={Giving feedback on others’ writing},
  author={Watling, Chris and Lingard, Lorelei},
  journal={Perspectives on medical education},
  volume={8},
  pages={25--27},
  year={2019},
  publisher={Springer}
}

@inproceedings{2023Petersen-ITiCSEWG,
  title={The robots are here: Navigating the generative ai revolution in computing education},
  author={Prather, James and Denny, Paul and Leinonen, Juho and Becker, Brett A and Albluwi, Ibrahim and Craig, Michelle and Keuning, Hieke and Kiesler, Natalie and Kohn, Tobias and Luxton-Reilly, Andrew and others},
  booktitle={Proceedings of the 2023 working group reports on innovation and technology in computer science education},
  pages={108--159},
  year={2023}
}

@inproceedings{10.1145/3649217.3653574,
  title={Desirable characteristics for ai teaching assistants in programming education},
  author={Denny, Paul and MacNeil, Stephen and Savelka, Jaromir and Porter, Leo and Luxton-Reilly, Andrew},
  booktitle={Proceedings of the 2024 on Innovation and Technology in Computer Science Education V. 1},
  pages={408--414},
  year={2024}
}

@inproceedings{10.1145/3626252.3630789,
  title={Beyond traditional teaching: Large language models as simulated teaching assistants in computer science},
  author={Liu, Mengqi and M'hiri, Faten},
  booktitle={Proceedings of the 55th ACM Technical Symposium on Computer Science Education V. 1},
  pages={743--749},
  year={2024}
}

@inproceedings{kazemitabaar2024codeaid,
  title={Codeaid: Evaluating a classroom deployment of an llm-based programming assistant that balances student and educator needs},
  author={Kazemitabaar, Majeed and Ye, Runlong and Wang, Xiaoning and Henley, Austin Zachary and Denny, Paul and Craig, Michelle and Grossman, Tovi},
  booktitle={Proceedings of the 2024 chi conference on human factors in computing systems},
  pages={1--20},
  year={2024}
}

@article{evans2013making,
  title={Making sense of assessment feedback in higher education},
  author={Evans, Carol},
  journal={Review of educational research},
  volume={83},
  number={1},
  pages={70--120},
  year={2013},
  publisher={Sage Publications Sage CA: Los Angeles, CA}
}

@article{bonsu2021influence,
  title={The influence of written feedback on the writing skill performance of high school students},
  author={Bonsu, Emmanuel},
  journal={International Journal of Applied Research in Social Sciences},
  volume={3},
  number={3},
  pages={33--43},
  year={2021}
}

@inproceedings{10.1145/3446871.3469768,
  title={Automated, Personalised, and Timely Feedback for Awareness of Programming Plagiarism and Collusion},
  author={Karnalim, Oscar},
  booktitle={Proceedings of the 17th ACM Conference on International Computing Education Research},
  pages={393--394},
  year={2021}
}

@article{suraworachet2023impact,
  title={Impact of combining human and analytics feedback on students’ engagement with, and performance in, reflective writing tasks},
  author={Suraworachet, Wannapon and Zhou, Qi and Cukurova, Mutlu},
  journal={International Journal of Educational Technology in Higher Education},
  volume={20},
  number={1},
  pages={1--24},
  year={2023},
  publisher={Springer}
}

@article{liu2024generating,
  title={Generating timely individualized feedback to support student learning of conceptual knowledge in Writing-To-Learn activities},
  author={Liu, Yang and Xiong, Wei and Xiong, Ye and Wu, Yi-fang Brook},
  journal={Journal of Computers in Education},
  volume={11},
  number={2},
  pages={367--399},
  year={2024},
  publisher={Springer}
}

@inproceedings{fisk2022automating,
  title={Automating Personalized Feedback to Improve Students' Persistence in Computing},
  author={Fisk, Susan and Hunt, Cynthia and Battestilli, Lina and Akram, Bita and Barnes, Tiffany and Price, Thomas and Yoder, Spencer},
  booktitle={Proceedings of the 53rd ACM Technical Symposium on Computer Science Education V. 2},
  pages={1197--1197},
  year={2022}
}

@article{nicol2006formative,
  title={Formative assessment and self-regulated learning: A model and seven principles of good feedback practice},
  author={Nicol, David J and Macfarlane-Dick, Debra},
  journal={Studies in higher education},
  volume={31},
  number={2},
  pages={199--218},
  year={2006},
  publisher={Taylor \& Francis}
}

@inproceedings{a,
  title={Integrating small language models with retrieval-augmented generation in computing education: Key takeaways, setup, and practical insights},
  author={Yu, Zezhu and Liu, Suqing and Denny, Paul and Bergen, Andreas and Liut, Michael},
  booktitle={Proceedings of the 56th ACM Technical Symposium on Computer Science Education V. 1},
  pages={1302--1308},
  year={2025}
}

@inproceedings{10.1145/3649217.3653554,
  title={Can small language models with retrieval-augmented generation replace large language models when learning computer science?},
  author={Liu, Suqing and Yu, Zezhu and Huang, Feiran and Bulbulia, Yousef and Bergen, Andreas and Liut, Michael},
  booktitle={Proceedings of the 2024 on Innovation and Technology in Computer Science Education V. 1},
  pages={388--393},
  year={2024}
}

@article{pădurean2025humanizingautomatedprogrammingfeedback,
  title={Humanizing Automated Programming Feedback: Fine-Tuning Generative Models with Student-Written Feedback},
  author={P{\u{a}}durean, Victor-Alexandru and Phung, Tung and Kotalwar, Nachiket and Liut, Michael and Leinonen, Juho and Denny, Paul and Singla, Adish},
  journal={arXiv preprint arXiv:2509.10647},
  year={2025}
}

@article{braun2006using,
  title={Using thematic analysis in psychology},
  author={Braun, Virginia and Clarke, Victoria},
  journal={Qualitative research in psychology},
  volume={3},
  number={2},
  pages={77--101},
  year={2006},
  publisher={Taylor \& Francis}
}

@article{hammer2014confusing,
  title={Confusing claims for data: A critique of common practices for presenting qualitative research on learning},
  author={Hammer, David and Berland, Leema K},
  journal={Journal of the Learning Sciences},
  volume={23},
  number={1},
  pages={37--46},
  year={2014},
  publisher={Taylor \& Francis}
}

@book{cohen2013statistical,
  title={Statistical power analysis for the behavioral sciences},
  author={Cohen, Jacob},
  year={2013},
  publisher={routledge}
}

@article{sweller1988cognitive,
  title={Cognitive load during problem solving: Effects on learning},
  author={Sweller, John},
  journal={Cognitive science},
  volume={12},
  number={2},
  pages={257--285},
  year={1988},
  publisher={Elsevier}
}

@article{sweller1998cognitive,
  title={Cognitive architecture and instructional design},
  author={Sweller, John and Van Merrienboer, Jeroen JG and Paas, Fred GWC},
  journal={Educational psychology review},
  volume={10},
  number={3},
  pages={251--296},
  year={1998},
  publisher={Springer}
}

@article{wood1976role,
  title={The role of tutoring in problem solving},
  author={Wood, David and Bruner, Jerome S and Ross, Gail},
  journal={Journal of child psychology and psychiatry},
  volume={17},
  number={2},
  pages={89--100},
  year={1976},
  publisher={Blackwell Publishing Ltd Oxford, UK}
}

@incollection{kalyuga2009expertise,
  title={The expertise reversal effect},
  author={Kalyuga, Slava},
  booktitle={Managing cognitive load in adaptive multimedia learning},
  pages={58--80},
  year={2009},
  publisher={IGI Global Scientific Publishing}
}

@article{grattafiori2024llama,
  title={The llama 3 herd of models},
  author={Grattafiori, Aaron and Dubey, Abhimanyu and Jauhri, Abhinav and Pandey, Abhinav and Kadian, Abhishek and Al-Dahle, Ahmad and Letman, Aiesha and Mathur, Akhil and Schelten, Alan and Vaughan, Alex and others},
  journal={arXiv preprint arXiv:2407.21783},
  year={2024}
}

@article{guo2024resist,
  title={To resist it or to embrace it? Examining ChatGPT’s potential to support teacher feedback in EFL writing},
  author={Guo, Kai and Wang, Deliang},
  journal={Education and Information Technologies},
  volume={29},
  number={7},
  pages={8435--8463},
  year={2024},
  publisher={Springer}
}

\newpage
\section*{Appendix}
\label{sec:appendix}

\subsection*{SLM Technical Specifications}
\label{appendix:tech-spec}
The SLM was locally hosted on a high-end Ubuntu workstation with an AMD Ryzen Threadripper PRO 7965WX 24-core/48-thread CPU, paired with 192GB of 6400MHz DDR5 memory, a 4TB Samsung 990 Pro NVMe SSD, and a GIGABYTE AORUS GeForce RTX 5090 GPU.

\begin{samepage}
\subsection*{LLM/SLM System Prompt}
\label{appendix:sys-prompt}

\begin{tikzpicture}
    \node[
        draw,
        rounded corners=5pt,
        thick,
        text width=\columnwidth,
        align=left,
        font=\small
    ] (gpt4) {
        You are an experienced writing professor teaching a course on [COURSE TOPIC].
        \\
        Students were required to complete the following assignment: [ASSIGNMENT INSTRUCTIONS].
        \\
        Your student submitted the following work for your feedback: [SUBMISSION].
        \\
        You are required to provide constructive feedback and meaningful steps for improvement
        based on the ``criteria'' and ``description'' outlined in the rubric. 
        \\
        Your feedback must 
        address each of the criteria from the rubric and must not assign a numerical grade.
        \\
        For each criterion, provide the following information: \\
        \hspace{2em} 1. The criteria name \\
        \hspace{2em} 2. Feedback on the student's performance \\
        \hspace{2em} 3. Reasoning behind the feedback \\
        \hspace{2em} 4. Specific suggestions for improvement \\

        Grading rubric: [Grading rubric of given course]

        Structure your feedback using the following format: \\
        \hspace{2em} Criteria: [Name of criteria] \\
        \hspace{2em} Feedback: [Detailed feedback] \\
        \hspace{2em} Reasoning: [Why this feedback was provided] \\
        \hspace{2em} Suggestions for Improvement: [Steps to improve on this criterion] \\
    };
\end{tikzpicture}
\end{samepage}

\end{document}